\documentstyle[aaspp4]{article}
\def\gsim{\;\rlap{\lower 2.5pt
 \hbox{$\sim$}}\raise 1.5pt\hbox{$>$}\;}
\def\lsim{\;\rlap{\lower 2.5pt
   \hbox{$\sim$}}\raise 1.5pt\hbox{$<$}\;}

\lefthead{HAIMAN \& MENOU}
\righthead{QUASAR EVOLUTION}

\begin{document}
\title{On the Cosmological Evolution of the Luminosity Function
and the Accretion Rate of Quasars}
\author{Zoltan Haiman}
\affil{NASA/Fermilab Astrophysics Center\\ Fermi National Accelerator
Laboratory, Batavia, IL 60510, USA, email: zoltan@fnal.gov}
\and
\author{Kristen Menou \altaffilmark{1}}
\affil{Harvard-Smithsonian Center for Astrophysics\\
60 Garden St., Cambridge, MA 02138, USA, email: kmenou@cfa.harvard.edu}

\altaffiltext{1}{also DARC-CNRS, Observatoire de Paris-Meudon,
92195 Meudon, C\'edex, France.}

\begin{abstract}
  We consider a class of models for the redshift evolution (between $0\lsim z
  \lsim 4$) of the observed optical and X--ray quasar luminosity functions
  (LFs), with the following assumptions: (i) the mass--function of dark matter
  halos follows the Press--Schechter theory, (ii) the black hole (BH) mass
  scales linearly with the halo mass, (iii) quasars have a constant universal
  lifetime, and (iv) a thin accretion disk provides the optical luminosity of
  quasars, while the X--ray/optical flux ratio is calibrated from a sample of
  observed quasars. The mass accretion rate $\dot{M}$ onto quasar BHs is a free
  parameter of the models, that we constrain using the observed LFs.  The
  accretion rate $\dot M$ inferred from either the optical or X--ray data under
  these assumptions generally decreases as a function of cosmic time from $z
  \simeq 4$ to $z \simeq 0$. 
  We find that a comparable accretion rate is inferred from the X--ray and 
  optical LF only if the X--ray/optical flux ratio decreases with BH mass.
  Near $z\simeq 0$, $\dot M$ drops to substantially
  sub-Eddington values at which advection--dominated accretion flows (ADAFs)
  exist.  Such a decline of $\dot M$, possibly followed by a transition to
  radiatively inefficient ADAFs, could explain both the absence of bright
  quasars in the local universe and the faintness of accreting BHs at the
  centers of nearby galaxies.  We argue that a decline of the accretion rate of
  the quasar population is indeed expected in cosmological structure formation
  models.
\end{abstract}


\section{Introduction}

The population of quasars as a whole exhibits a characteristic cosmological
evolution: the number density of quasars rises monotonically by two orders of
magnitude from redshift $z\simeq 0.1$ to an apparent peak at $z_{\rm
pk}\simeq2.5$. The evolution at redshifts exceeding $z_{\rm pk}$ is still
unclear: the number density of optically bright quasars declines from $z_{\rm
pk}\simeq2.5$ to $z\simeq 4.5$ (Pei 1995), but recent ROSAT data have not shown
any evidence for a similar decline in X--rays (Miyaji et al.  1998). The
accreting supermassive black hole model (first proposed by Salpeter (1964) and
Zel'dovich (1964); see also Rees 1984) produces the bolometric luminosity of an
individual quasar, and provides a framework that has successfully accounted for
several properties of the quasar phenomenology (e.g. Blandford 1992).  However,
the reason behind the evolution of the quasar luminosity function must involve
some additional physics, likely related to the cosmological growth of
structures (Efstathiou \& Rees 1988)

Several ideas have been put forward to explain quasar evolution, including
activation by mergers (Carlberg 1990); intermittent accretion (Small \&
Blandford 1992); association with dark halos using a nonlinear (Haehnelt \&
Rees 1993) or linear (Haiman \& Loeb 1998) scaling of the BH mass $M_{\rm bh}$
with halo mass $M_{\rm halo}$; a relation to the individual quasar light curves
(Siemiginowska \& Elvis 1997) or spectral shapes (Caditz et al. 1991); and a
transition to ADAFs at low redshift (Yi 1996).  More recently, Cavaliere \&
Vittorini (1998) have attributed the evolution to the initial assembly of the
host galaxies ($z\gsim 3$), followed by their intermittent interactions as
members of groups ($z\lsim 3$). All these ideas are still poorly constrained by
present data, and a conclusive understanding has not yet been achieved.

Quasars are virtually absent from the local universe, but the presence
of supermassive black holes (``dead quasars'') in inactive galaxies is
expected as a result of past quasar activity (e.g. Lynden--Bell 1969;
Soltan 1982; Rees 1990).  In recent years, significant progress has
been made in the detection of such supermassive black holes (see
Kormendy \& Richstone 1995 for a review), and the latest results
suggest that the nuclei of most, if not all, nearby galaxies contain
massive dark objects, presumably the remnants of quasars (Magorrian et
al. 1998). Fabian \& Canizares (1988) pointed out that the nuclei of
nearby elliptical galaxies appear much too dim to contain accreting
supermassive black holes given the estimated mass accretion rates.
Fabian \& Rees (1995) suggested that this could be due to accretion
via a radiatively inefficient ``ion torus'' (similar to an ADAF
for that matter; see Ichimaru 1977; Rees et al.  1982; Narayan \& Yi
1994; Abramowicz et al. 1995; or Narayan et al. 1998 for a review)
which exist only at sub-Eddington accretion rates. This idea was later
strengthened by calculations of Mahadevan (1997), and by the relative
success of several ADAF spectral models of specific low luminosity
nuclei (e.g. Lasota et al.  1996; Reynolds et al. 1996; Di Matteo et
al.  1996; see also Di Matteo et al. 1999).

The main ingredients in modeling the quasar LF is the BH formation rate and the
light--curve of each BH in the relevant wavelength band.  The usual approach is
to specify these ingredients with several parameters, and fit the LF by a
``trial and error'' procedure.  We emphasize here that the least understood
ingredient in these attempts is the mass accretion rate onto quasar black
holes.  Indeed, the key to understanding the cosmic evolution of quasars
probably lies in their accretion history.  In this paper, we invert the
problem, and infer the accretion rates of quasars directly from the observed
LF, while using a set of reasonable (although not unique) assumptions for the
other model ingredients.  Under these assumptions, we infer an accretion
rate for the quasar population, which generally decreases as a
function of time. By construction, using the inferred accretion rate as an
'ansatz' then yields a  fit to the evolution of the quasar LF. We find that
this fit is a
viable alternative to the models proposed so far, which assume an arbitrary
(often constant) accretion rate, and propose a redshift dependence for other
model parameters. An attractive feature of the class of models proposed here
is that the decline of accretion rates over cosmic time--scales could provide
a natural explanation for both the disappearance of bright quasars in the
local universe and the presence of faint, accreting supermassive black holes
at the center of most nearby galaxies.

Although we focus on optical and X-ray data, the method to infer $\dot M$
described here is applicable to any wavelength at which a LF and a reliable
emission model exist.  In this paper we adopt the concordance cosmology of
Ostriker \& Steinhardt (1995), i.e.\ a flat $\Lambda$CDM model with a slightly
tilted power spectrum ($\Omega_0,\Omega_\Lambda, \Omega_{\rm
b},h,\sigma_{8h^{-1}},n$)=(0.35, 0.65, 0.04, 0.65, 0.87, 0.96).  We have
verified that our conclusions below regarding the decline in the accretion rate
do not change significantly in other (e.g. open CDM) cosmologies.

The rest of this paper is organized as follows. In \S~2, we state our
assumptions on the masses of BHs; in \S~3, we discuss the procedure to
relate each quasar luminosity to its host halo's mass; and in \S~4, we
summarize the main features of the disk emission model used.  In \S~5,
we derive the accretion rate as a function of redshift and halo mass,
which is the main result of this paper.  In \S~6, we argue that a
declining accretion rate is expected in cosmological structure
formation models.  In \S~7, we address the question of an intrinsic
scatter around the mean ratio of BH to halo mass, and argue that such
a scatter would not significantly change our conclusions. In \S~8, we
briefly discuss the implication of the X--ray luminosity
function. Finally, in
\S~9 we discuss the implications of this work, and in \S~10 we
summarize our main conclusions.

\section{The Formation of Quasar Black Holes}

Although there is no a--priori theory for the cosmic black hole formation rate,
it is natural to expect that it is related to the formation of dark matter
halos. Although the fueling of the central BH takes place a the inner
few parsecs of the halo, both the central gas density and temperature, likely
to be the parameters controlling the fueling rate, are strongly
correlated with the depth of the initial dark matter potential well, and
therefore with the halo mass.  The tight correlation discovered between bulge
mass and black hole mass in nearby galaxies (Magorrian et al. 1998)
provides empirical evidence that two very different scales
(i.e. the inner few parsecs for the fueling, and the $\sim$kpc sized bulge)
are connected to each other.
The mass function of halos $dN_{\rm ps}/dM$ at any redshift is
described by the Press--Schechter (1974) theory with an accuracy of $\sim50\%$
when compared to numerical simulations at the mass--scales relevant here
(Somerville et al.  1998).  Note that the shape of the halo mass function has
the desirable property of a steep decline with $M_{\rm halo}$, similar to the
quasar LF.
It has been a tradition in QSO evolution studies
to phenomenologically characterize the LF evolution either as
pure luminosity evolution (PLE) or as pure number density
evolution (PDE).  A distinction between these two possibilities
was not feasible with the early data on the LF, which was consistent
with a simple power--law, lacking any features.  More recent data
(see e.g. Pei 1995 and references therein) have revealed a break
(or "knee") in the LF.
Although the redshift-evolution can still be adequately characterized as
PLE or PDE in the intermediate redshift range ($0.7\lsim z \lsim 1.7$;
see Hatziminaoglou et al. 1998), it appears that over the extended
interval of $0\lsim z \lsim 5$ the evolution of the LF requires a
more complex description
(Caditz \& Petrosian 1990; Pei 1995; Goldschmidt \& Miller 1997;
Miyaji et al. 1998). Associating BHs with dark matter
halos in the Press--Schechter
theory naturally gives rise to a mixture of luminosity and number
density evolutions.

Recent measurements of the black hole masses in 36 nearby galaxies
provided evidence for a linear relation $M_{\rm bh}\propto M_{\rm bulge}$
between black hole and bulge mass (Magorrian et al. 1998).  Assuming $M_{\rm
bulge}\propto M_{\rm halo}$, this result further implies $M_{\rm bh}\propto
M_{\rm halo}$.  A simple scenario in which the latter relation is satisfied in
the local universe is if the relation holds at every redshift, i.e.  if the BH
masses grow synchronously with the host halo masses, $\dot M_{\rm bh}/M_{\rm
bh} = \dot M_{\rm halo}/M_{\rm halo} \Rightarrow M_{\rm bh}/M_{\rm halo}={\rm
const}$.
Note that if mergers are important (i.e. if they occur on a
time--scale less than the Hubble time), then we must also suppose that black
holes merge together whenever their halos do.  In what follows, we assume that
$M_{\rm bh}/M_{\rm halo}=10^{-3.2}$, consistent with the value $M_{\rm
bh}/M_{\rm bulge}=0.006$ measured by Magorrian et al. (1998) if all the
baryonic mass is contained in the bulges of galaxy-size halos (see \S 7 for a
discussion of the effect of an intrinsic scatter in the ratio $M_{\rm
bh}/M_{\rm halo}$).
 We emphasize here that the correlation was observed by Magorrian et al.
(1998) in nearby galaxies, and therefore there is no observational
justification for our simplifying assumption that the same correlation
is already present as an "initial condition" at high redshifts.
A possibility that can not be ruled out is that this correlation is
established during the complex evolutionary histories of the individual
black holes.  This scenario would require significantly more complicated
modeling; however, the general feature of the models proposed here, namely a
gradually declining fueling rate, would likely remain valid. In particular, to
avoid a declining accretion rate, the ratio $M_{\rm bh}/M_{\rm halo}$ would
have to be $\sim$2 orders of magnitude higher at high redshift ($z\sim4$) than
at $z\approx 0$, which would imply that essentially all of the gas inside
high-$z$ halos was already incorporated into their central BHs.

\section{The Observed Quasar Luminosity Functions}

The quasar LF between redshifts $0.1\lsim z \lsim 4.5$ has been measured in the
optical; we use the fitting functions for the K--corrected rest--frame B--band
(around $\nu=10^{14.83}$Hz) LF obtained by Pei (1995).  More recently, the LF
has also been measured in X--rays (0.5--2 keV); we use the parametric fits
constructed by Miyaji et al. (1998).  Both the optical and X--ray LFs have a
steep slope, and the abundance of quasars rises rapidly from $z\simeq 0.1$ to
$z_{\rm pk} \simeq 2.5$.  At redshifts exceeding $z_{\rm pk}$, the abundance of
optically bright quasars turns over and slowly declines, while the X--ray data
reveal a LF that stays flat up to $z\simeq 4.5$.

The observed LF, $dN_{\rm obs}/dL$ can be compared directly with the
Press--Schechter halo mass function $dN_{\rm ps}/dM$ to derive the
luminosity $L(z,M)$ of each halo of mass $M$ that satisfies
\begin{equation}
\frac{dN_{\rm obs}}{dL}(z,L)\times\frac{dL}{dM}=f_{\rm on}
\frac{dN_{\rm ps}}{dM}(z,M).
\end{equation}
\noindent

\noindent
Here $f_{\rm on}$ is the quasar ``duty cycle'', i.e. the fraction of BHs
(assuming all halos contain BHs, see Magorrian et al. 1998) that are active at
a given time around redshift $z$.  This fraction is determined by the details
of quasar light--curves and the distribution of formation times of BHs, and
could be a function of several parameters, such as redshift, halo mass, etc.
Here we simply assume that each quasar shines for a constant time $t_{\rm
source}$, and we take the duty cycle to be $f_{\rm on}=t_{\rm source}/t_{\rm
Hub}(z)$, where $t_{\rm Hub}(z)$ is the redshift--dependent Hubble time.  Our
assumption is justified if the light--curve typically drops sharply a time
$t_{\rm source}$ after the quasar becomes active (e.g. due to exhaustion of
fuel, or shut--off due to a merger), or if intermittent activity adds up to a
total duration $t_{\rm source}$ (e.g. if fueling is due to discrete clumps of
gas). Note that $f_{\rm on}$ must be $\ll 1$ to avoid unrealistically large
masses of the quasar host halos, independently of the ratio $M_{\rm bh}/M_{\rm
halo}$.

\begin{figure}[t]
\includegraphics{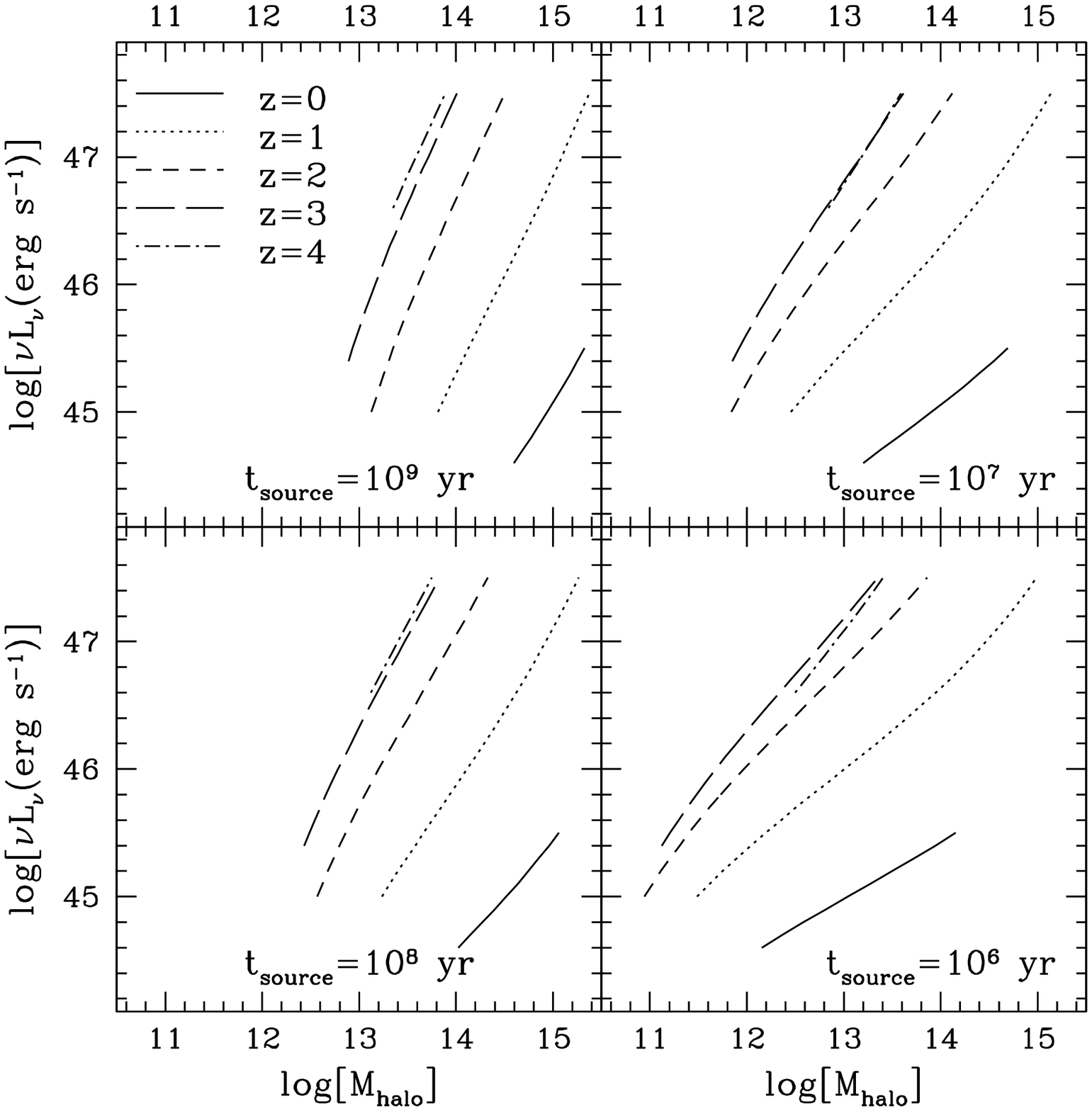}
\vspace*{4.5in}
\caption{The relation between quasar luminosity (in the
rest--frame B--band at $\nu_{\rm B}=10^{14.83}~{\rm Hz}$) and total halo mass,
derived by equating the space densities of the optical LF (Pei 1995) and the
Press--Schechter mass function. The relation is shown for 4 different quasar
lifetimes, at various redshifts.}
\label{fig:LM}
\end{figure}

The B--band luminosity $L_{\rm B}(M,z)$ associated with halo mass $M$ at
redshift $z$ obtained by integrating equation (1) is shown in
Figure~\ref{fig:LM} for $z=0,1,2,3$ and 4, and duty cycles corresponding to
$10^{6,7,8,9}$ yrs.  For a reasonable range of $t_{\rm source}$, observed
quasars are associated with halos of mass $\sim 10^{11}-10^{15}{\rm
M_{\odot}}$.  Note that this conclusion is based only on equating the observed
and predicted space densities; the only assumption in Figure~\ref{fig:LM}
(apart from adopting the Press--Schechter theory and the cosmological
parameters) is the constancy of the quasar lifetime. The general trend
suggested by Figure~\ref{fig:LM} is that, as time goes on, observed quasars are
less and less luminous, but correspond to more and more massive halos.

\section{Accretion Disk Models}

The optical/UV ``Big Blue Bump'' feature in the spectra of quasars is often
interpreted as thermal emission from a thin accretion disk around a
supermassive black hole (e.g.  Malkan 1983; Laor 1990; see also Collin-Souffrin
1994 and Antonucci 1999 for possible difficulties with this
interpretation). Here, we assume that, at any redshift, (i) the quasar
luminosity in the B--band originates from a steady thin disk (Shakura \&
Sunyaev 1973; Frank et al. 1992) around a non--rotating BH, (ii) the disk is
inclined at $i=60^o$ from the line of sight and (iii) general relativistic
effects, irradiation effects and possible complications due to
radiation-pressure dominated zones in the disk can be neglected.

Figure~\ref{fig:disk} shows the luminosities $L_{\rm disk}$ predicted by these
disk models. The lines show $L_{\rm disk}$ for various accretion rates ($\log
\dot m = 0$ to $-3$ with $0.5$ intervals, where $\dot m\equiv\dot M /\dot
M_{\rm Edd}$) as a function of the central BH mass $M_{\rm BH}$. Note
that here and in the remainder of this paper, we use the Eddington
accretion rate $\dot M_{\rm Edd}=1.4\times 10^{18} (M_{\rm
BH}/M_{\odot})~{\rm g~s^{-1}}$ corresponding to a $10\%$ radiative
efficiency.

The essential feature of the disk emission used in our work is that
there is a unique $L_{\rm disk}$ predicted in the B--band for a given
$M_{\rm BH}$ and $\dot m$, as seen from Figure~\ref{fig:disk}.  The
reduction of $L_{\rm disk}$ at high $M_{\rm BH}$ and low $\dot m$ in
this figure appears because for this range of parameters the disk
emission peaks at wavelengths longer than the B--band (see also Caditz
et al. 1991).  This characteristic feature of the spectrum
affects our conclusions for the accretion rates only for the most
massive BHs at low redshifts, where the accretion rate is found to be
substantially sub--Eddington.  If this effect was omitted from our
calculations, the accretion rate for a given quasar luminosity at low
redshifts would be smaller than the values inferred in \S~5.  We are
therefore conservative in including this effect.

\section{Evolution of the Quasar Accretion Rate}

\begin{figure}[t]
\includegraphics{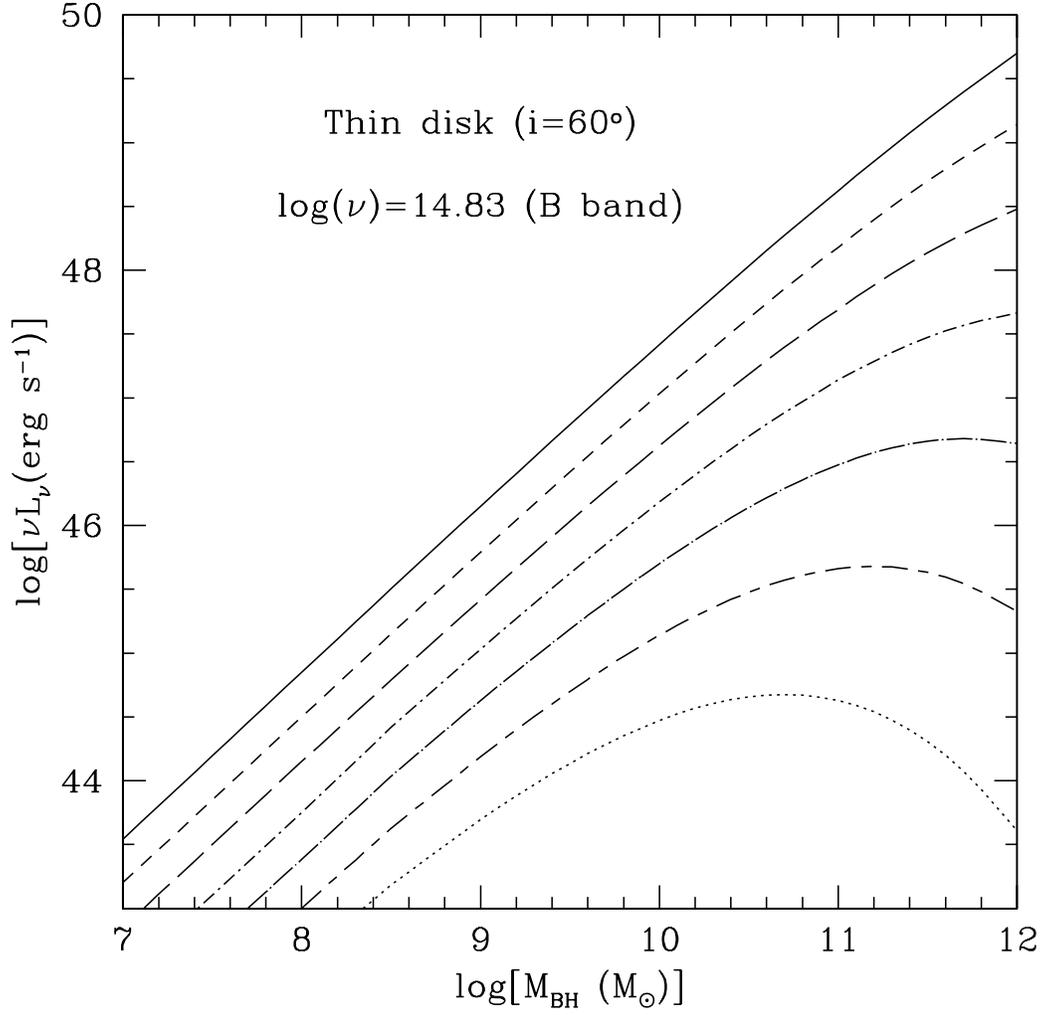}
\vspace*{4.5in}
\caption{Predictions for the rest-frame B--band luminosity of an accretion disk
inclined at $i=60^o$ from the line of sight. The lines show the luminosities
predicted for various accretion rates (from top to bottom, $\log~\dot m=0$ to
$-3$ with $0.5$ intervals in Eddington units) as a function of the central
black hole mass $M_{\rm bh}$.}
\label{fig:disk}
\end{figure}

The accretion rate in quasars as a function of redshift and BH mass can be
found by equating the observed and predicted luminosities, $L_{\rm B}(M_{\rm
bh},z)=L_{\rm disk}(M_{\rm bh},\dot m)$.  Here $L_{\rm B}$ is the B band
luminosity associated with a halo of mass $M_{\rm halo}$ (Fig. 1), which
harbors a black hole of mass $M_{\rm bh}=10^{-3.2}M_{\rm halo}$; $L_{\rm disk}$
is the B--band luminosity of the quasar as shown in Figure~\ref{fig:disk}. As
an example, according to Figure~\ref{fig:LM} if the source lifetime is $t_{\rm
source}=10^7$ yr, at $z=2$ a halo of mass $M_{\rm halo}=10^{12.2}~{\rm
M_\odot}$ (with a black hole of mass $M_{\rm bh}=10^9~{\rm M_\odot}$) has a
luminosity of $10^{45.5}~{\rm erg~s^{-1}}$. This luminosity and black hole
mass, according to Figure~\ref{fig:disk}, correspond to an accretion rate $\dot
m = 0.1$ (third curve from top).

\begin{figure}[t]
\includegraphics{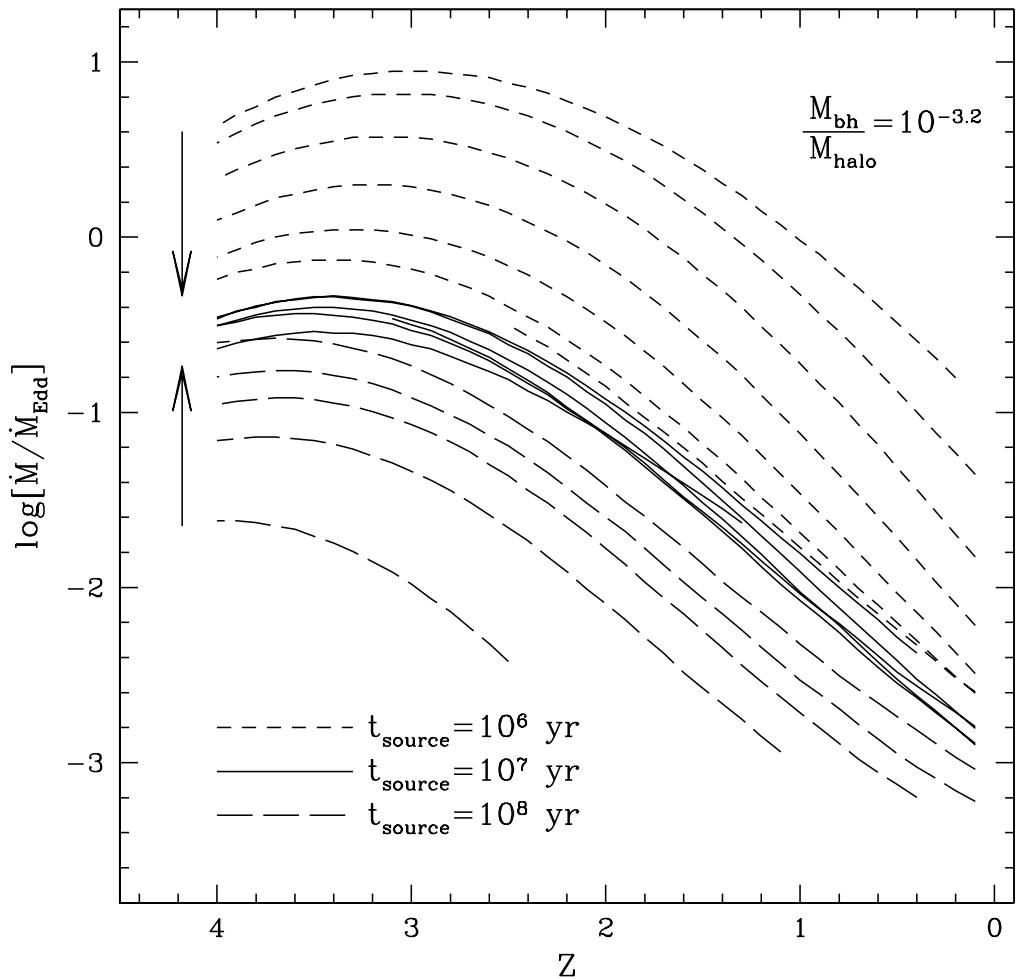}
\vspace*{4.5in}
\caption{Accretion rates inferred from the quasar
  B--band LF and Press-Schechter theory, assuming a constant ratio $M_{\rm
  bh}/M_{\rm halo}=10^{-3.2}$. The various curves correspond to halo masses in
  the range $10^{11} \leq M_{\rm halo}/{\rm M_\odot} \leq 10^{14}$, and the
  results are shown for three different values of the quasar lifetime.  Note
  that the accretion rates are shown for halos with a fixed mass $M_{\rm
  halo}$, rather than following the accretion rate of a single halo with a
  given initial mass. The arrows indicate the direction of increasing halo mass
  for the two cases $t_{\rm source}=10^6$ and $10^8$ yr.}
\label{fig:mdot}
\end{figure}

In Figure~\ref{fig:mdot}, we show the inferred accretion rate $\dot m$ as a
function of redshift for three different duty cycles ($t_{\rm
source}=10^{6,7,8}$ yr).  In each case, we show the accretion rates for the
halo mass range $10^{11} \leq M_{\rm halo}/{\rm M_\odot} \leq 10^{14}$.
Figure~\ref{fig:mdot} reveals that in all cases, the accretion rate decreases
with decreasing redshift, and that between $3\gsim z \gsim 0$, it drops by a
factor of several hundred.

The inferred accretion rates are roughly independent of halo mass for $t_{\rm
source}=10^7$ yr, as reflected by the grouping of the solid lines in
Figure~\ref{fig:mdot}.  This grouping is partly due to the combined effect of
the non--linear dependence of $L_{\rm B}$ on M${\rm halo}$ (Fig.~\ref{fig:LM}),
and the turnover in the luminosity at high BH masses and low accretion rates
(Fig.~\ref{fig:disk}).  For a longer lifetime, $\dot m$ increases with
increasing halo mass (long dashed lines); while for a shorter lifetime $\dot m$
decreases with increasing halo mass (short dashed lines).  Note that in the
latter case $\dot m>1$ is predicted, meaning that the required luminosities can
be produced only by BHs larger than we assumed; this would be inconsistent with
the observational results of Magorrian et al. (1998).  Although
Figure~\ref{fig:mdot} shows results only from the optical LF, we obtain
qualitatively similar curves from the X--ray data, if we assume an
X--ray/optical flux ratio of $\sim20\%$.  This value is based on the mean
spectrum of a sample of 48 quasars, selected by Elvis et al. (1994) by the
requirement that each quasar has a good spectrum both in the optical (from {\it
IUE}) and X-ray (from {\it Einstein}).

The evolution of $\dot m$, independent of $M_{\rm halo}$ for $t_{\rm
source}=10^7$ yr, suggests that the accretion rate in physical units goes as
$\dot M \propto M$ at any given time.  Such a scaling would hold, for example,
if each halo was undergoing isolated spherical self-similar infall
(Bertschinger 1985).  The apparent turnover of $\dot m$ at redshifts $z\gsim 3$
reflects the decline of the B--band LF at these redshifts, and may be caused by
dust obscuration (Heisler \& Ostriker 1988).  This explanation is consistent
with the recent X--ray data (insensitive to the presence or absence of dust)
showing that the X-ray LF does not decline at $z\gsim 3$ (see \S~8 for a more
detailed comparison between X-ray and optical data).

\section{Accretion Rate From Structure Formation Models}

Is a declining accretion rate in the population of quasars expected?  A
derivation of the accretion rate from first principles must necessarily address
several complicated issues that are beyond the scope of this paper, such as the
role of angular momentum and gas cooling.  
Such studies are now underway, e.g. Sellwood \& Moore (1999) have
explored in detail the formation of massive central objects in bars of
galactic disks, and their subsequent fueling, including the dependence
on angular momentum.
Here we will only put forward two general arguments for a decline of 
the accretion rate with decreasing redshift.

\begin{figure}[t]
\includegraphics{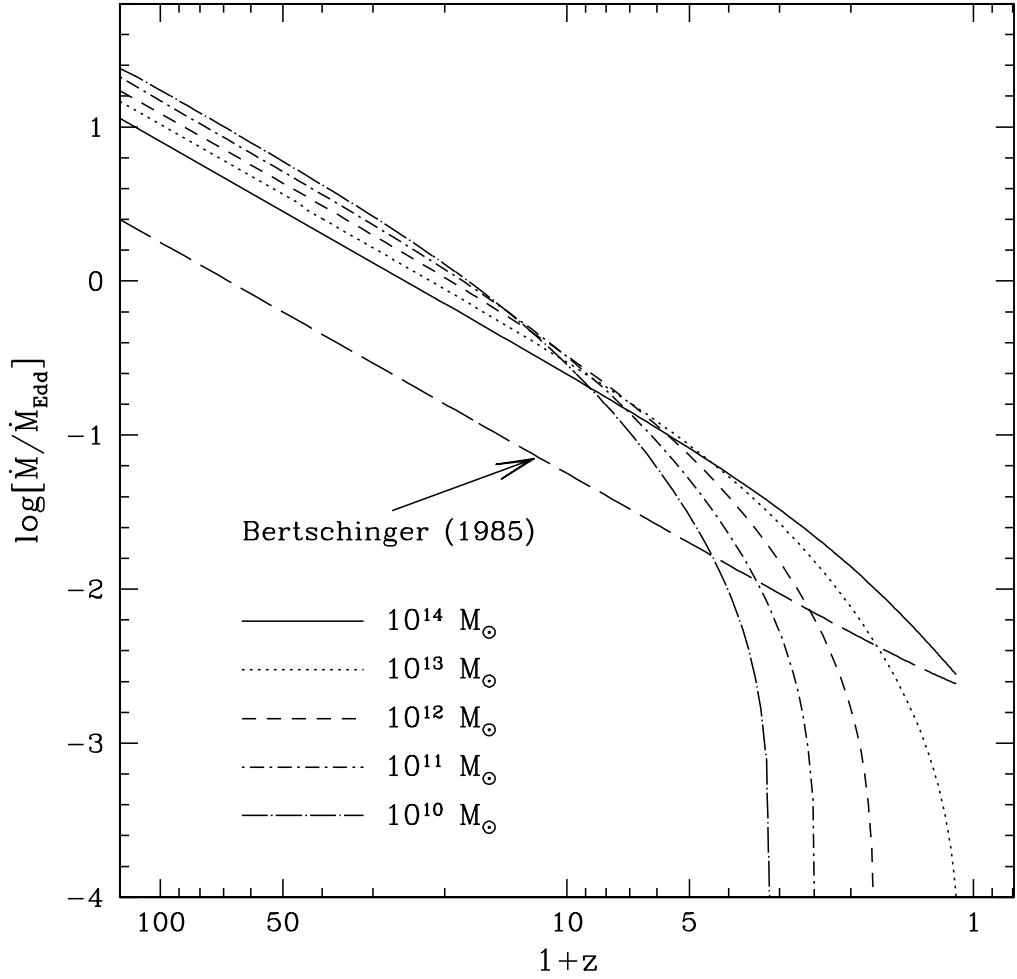}
\vspace*{4.5in}
\caption{Accretion rates predicted by Press--Schechter theory
for halos of various masses, assuming that the evolution of the PS mass
function is driven by accretion only. Also shown is the accretion rate
predicted by the self-similar collapse theory of Bertschinger (1985). The
accretion rates for the central BHs are identical to those of the halos if the
ratio $M_{\rm bh}/M_{\rm halo}$ is constant.}
\label{fig:mdot2}
\end{figure}

First, assuming that each black hole grows as a central point mass in an
isolated collapsing spherical cloud, the self--similar infall solutions of
Bertschinger (1985) imply\footnote{Although these relations are valid only for
an Einstein--de Sitter universe, a similar behavior is expected in
$\Omega+\Lambda$=1 models.}  that $M_{\rm bh} \propto t^{2/3}$ and
$\dot{M}_{\rm bh}/M_{\rm bh}=2/3t$.  Although isolated spherical infall is
likely to be a crude approximation at best for the formation of quasar black
holes, the primary reason for the slow growth of the central mass is the
cosmological expansion\footnote{For comparison, note that in a static medium
the central mass would grow exponentially or faster; see Bondi (1952).}.  It is
therefore unlikely that departure from spherical symmetry, or interactions with
other collapsing halos could considerably speed up the accretion, relative to
the power--law growth for an extended period of time (corresponding to $0\lsim
z\lsim 4$) in the entire quasar population.  In Figure~\ref{fig:mdot2} we show
the (declining) accretion rate as a function of redshift based on the
Bertschinger (1985) solutions.

A second argument can be drawn directly from the Press--Schechter theory.
In the limit that $d/dt (dN_{\rm ps}/dM)$ merely reflects the growth
in mass of each individual halo, the accretion rate of halos can be
obtained by requiring that the total number density of halos does not
change, $d/dt [M dN/dM] = 0$.  This yields for the halo mass accretion
rate
\begin{equation}
\dot m = - \frac{1}{M} \times
\frac{\partial}{\partial t} \left( \frac{dN_{\rm ps}}{dM} \right) \times \left[ \frac{1}{M} \left( \frac{dN_{\rm ps}}{dM} \right) +
\frac{\partial}{\partial M} \left( \frac{dN_{\rm ps}}{dM} \right) \right]^{-1},
\end{equation}
which, by our assumption of constant ratio $M_{\rm bh}/M_{\rm halo}$,
is identical to the BH accretion rate.  In Figure~\ref{fig:mdot2} we
show the accretion rates in this limit as a function of redshift.  The
accretion rates drop in all cases, but unlike in the isolated
spherical self-similar infall case, $\dot m(z)$ depends on the halo
mass.  Note that both the Bertschinger solution and Press--Schechter
theory predict mass accretion rates which are super--Eddington at high
$z$, and become substantially sub--Eddington near $z \sim 0$.

An important ingredient ignored in the above argument is the occurrence
of mergers between halos.  Naively taking a time--derivative of the
mass--function $dN_{\rm ps}/dM$ results in a quantity that is negative
for small masses, and positive for large masses, with the sign
changing at a critical value $M=M_*$.  Sasaki (1994) argued that this
behavior reflects a decreasing contribution of destructive mergers to
the evolution of the mass function for halos of mass $M>M_*$.  In
other words, the evolution of the mass function at large masses (the
typical halo mass $M_{\rm halo}$ of interest here) would be governed
mainly by accretion, rather than mergers.

\begin{figure}[t]
\includegraphics{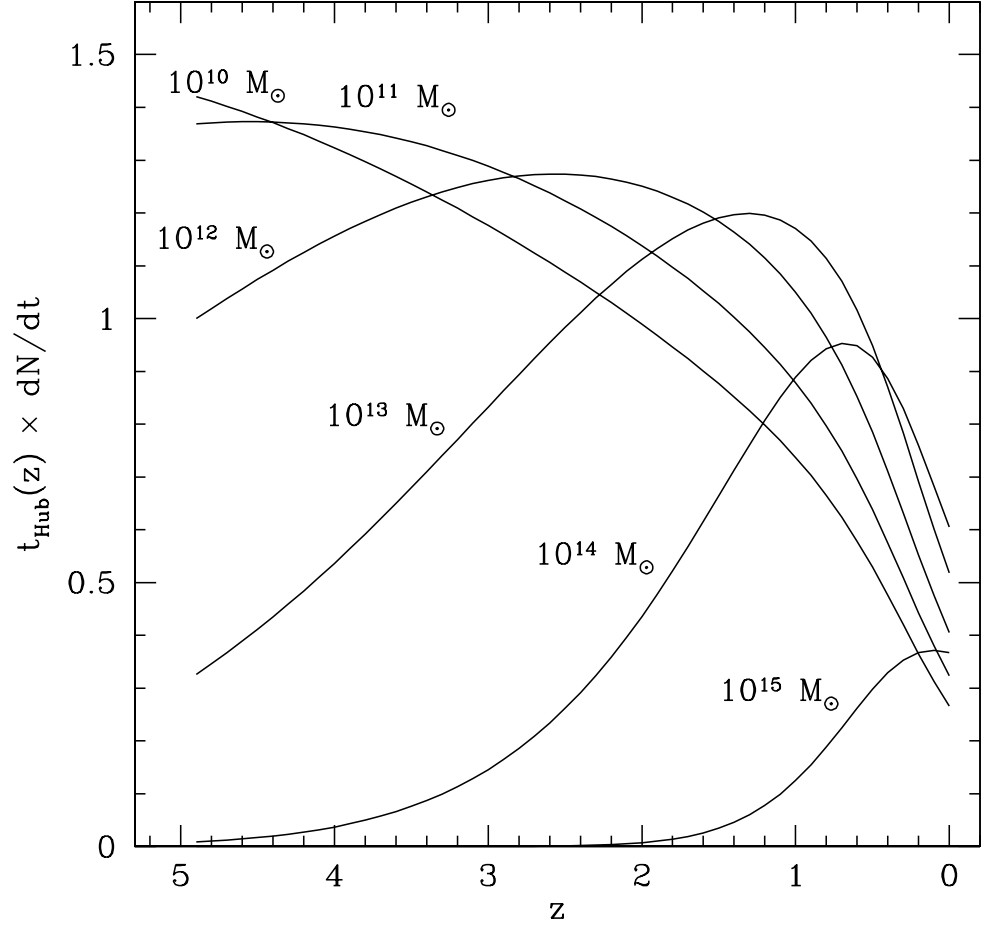}
\vspace*{4.5in}
\caption{The number of major mergers that a halo of a given mass experiences in
a Hubble time (see text for details), according to the extended Press-Schechter
formalism of Lacey \& Cole (1993) }
\label{fig:merge}
\end{figure}

In order to quantify the role of mergers, we have used the extended
Press--Schechter formalism (Lacey \& Cole 1993, equation 2.18) to
compute merger rates between halos of various masses.  In
figure~\ref{fig:merge} we show the number of major mergers that a halo
of a given mass $M_{\rm halo}=10^{10-15}~{\rm M_\odot}$ experiences in
a Hubble time.  We have defined a major merger to occur when the mass
ratio of the merging halos is less than 0.5 (i.e. a halo collides with
another halo whose mass is between $M_{\rm halo}/2$ and $2M_{\rm
  halo}$).  The figure reveals that the merger rate between
galaxy--sized halos $(\sim 10^{12}~{\rm M_\odot})$ has the desirable
property of having a well--defined peak around $z\sim2.5$, the
redshift at which the LF of bright quasars peaks.  However, the figure
also shows that the decrease in the physical merger rate $dN/dt$
between $z=3$ and $z=0$ is only a factor of $\sim 15$ (taking into
account a factor of $\sim 5$ increase in the Hubble time during this
interval).  If the fueling rate of BHs were proportional to the merger
rate (corresponding to a fixed fraction of the baryons going into the
hole for each merger), this decrease would be insufficient to explain
the evolution of the observed quasar LF: our results show that the
decrease in $\dot m$ has to be over two orders of magnitude.  Note
that our definition of a ``major merger'', and therefore the resulting
merger rate, is somewhat arbitrary. However, we find that including a
wider range of halo masses in defining the merger rate still leads to
an insufficiently small decrease in $\dot m$.

We conclude that the decrease in the merger rates between halos might
play an important role in reducing the accretion rates, and contribute
to the cosmological evolution of the quasar LF.  However, we find that
if the fueling rate of quasars is proportional to the halo merger rate,
then the rapid evolution between $z=3$ and $z=0$ of the quasar LF
requires an additional decrease by about an order of magnitude in the
fueling rate.

\section{The Role of Intrinsic Scatter}

We have so far assumed a deterministic linear relation between the quasar
luminosity $L$, halo mass $M_{\rm halo}$ and black hole mass $M_{\rm bh}$.
More realistically, one would expect that individual quasars exhibit an
intrinsic scatter around the mean ratio $M_{\rm bh}/M_{\rm halo}$.  In the
sample of 36 galaxies analyzed by Magorrian et al. (1998), the value of
$r=M_{\rm bh}/M_{\rm bulge}$ is found to obey the probability distribution
$p(r)=\exp [-(\log r - \log r_0)^2/2\sigma^2]$ with $\log r_0=-2.28$ and
$\sigma=0.51$.  The implications of this result for the scatter in the ratio
$x=M_{\rm bh}/M_{\rm halo}$ is not clear, because of our ignorance of $M_{\rm
halo}$ for the observed galaxies.  Van der Marel (1999) finds statistical
errors in the observationally determined BH masses (due to photometry, PSF
deconvolution, and uncertainties about stellar contributions to the central
density profile) that could explain all of the observed scatter.  However, some
unknown fraction of the observed scatter $\sigma\sim0.5$ is likely to be real,
and we investigate below the effect of this scatter on the inferred accretion
rates.

In the presence of an intrinsic scatter, the BH mass function is given by the
convolution $dN/dM_{\rm BH}=\int dx (1/x) p(x) dN/dM_{\rm halo}(M_{\rm bh}/x)$.
At large halo masses, where the halo mass function is a steeply declining
function of $M_{\rm halo}$, this convolution increases the expected space
density of BHs of a given mass.  To quantify the effect that this increase has
on the models, we assume the ``log--normal'' probability distribution given
above, with $\log x_0=-3.2$ and $\sigma=0.5$ for the ratio $x=M_{\rm bh}/M_{\rm
halo}$.  This is equivalent to the extreme assumption that {\it all} of the
observed scatter is intrinsic; the real situation is likely to be in between
the deterministic linear relation (\S~5) and the case of strong intrinsic
scatter presented here.

\begin{figure}[t]
\includegraphics{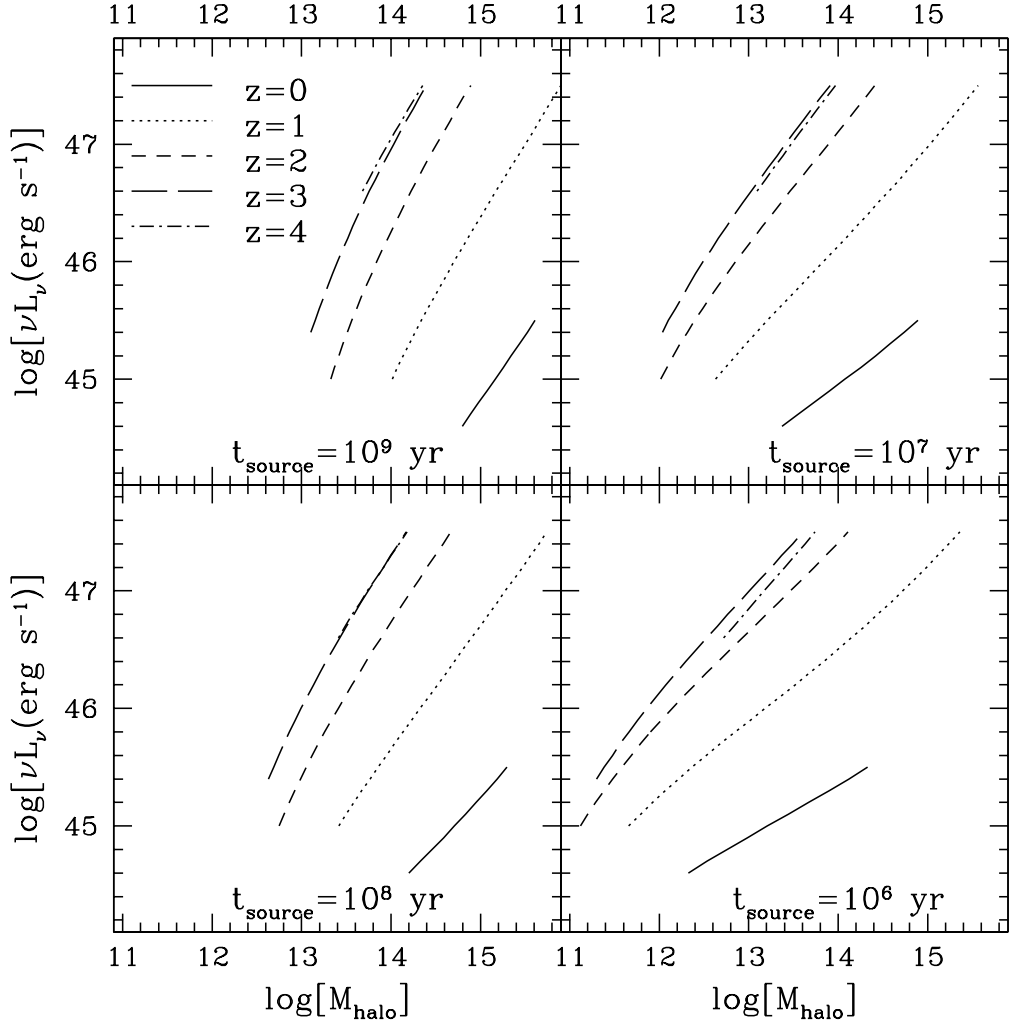}
\vspace*{4.5in}
\caption{The mean relation between quasar luminosity (in the
rest--frame B--band at $\nu_{\rm B}=10^{14.83}~{\rm Hz}$) and total halo mass
in the presence of intrinsic scatter of half an order of magnitude (up and
down) in the ration $M_{\rm bh}/M_{\rm halo}$ (compare to Figure~1).}
\label{fig:LMscat}
\end{figure}

\begin{figure}[t]
\includegraphics{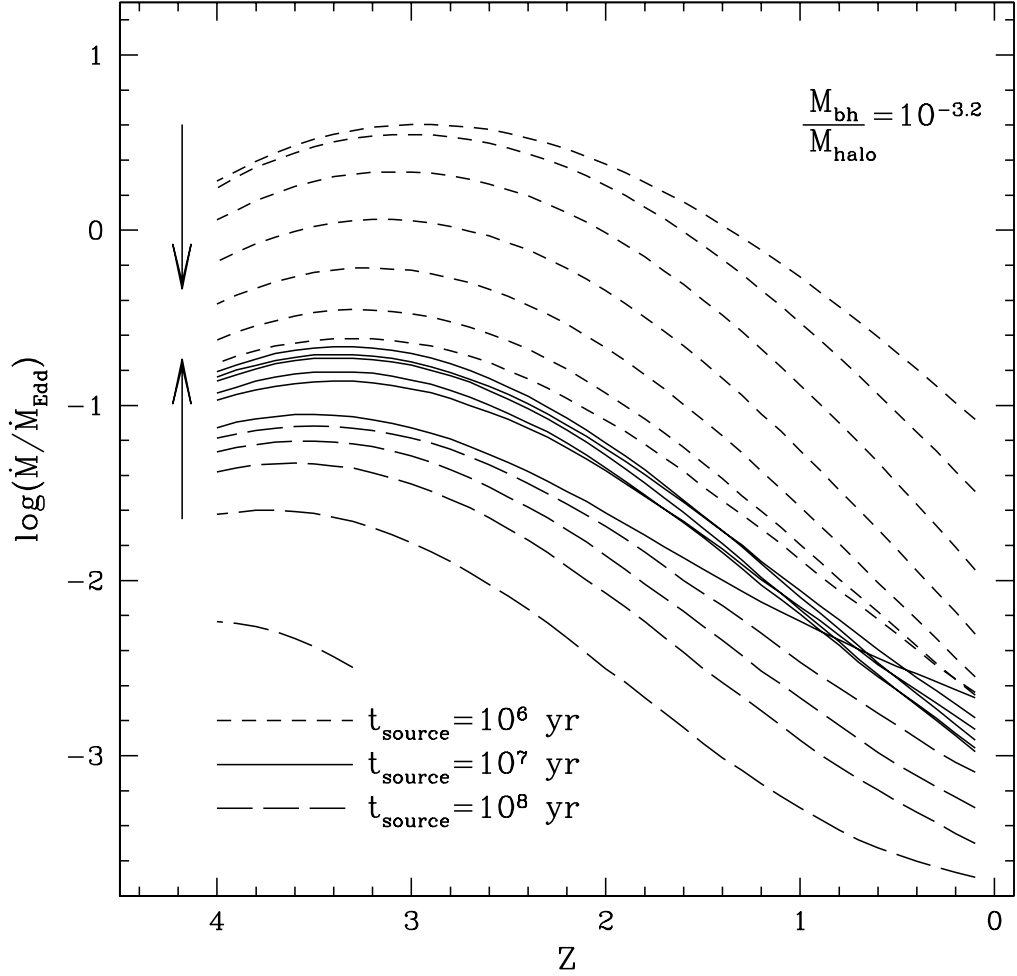}
\vspace*{4.5in}
\caption{Mean accretion rates inferred from the quasar optical LF and
Press-Schechter theory, assuming a {\it mean} ratio $M_{\rm bh}/M_{\rm
halo}=10^{-3.2}$ with an intrinsic scatter in $\log (M_{\rm bh}/M_{\rm halo})$
of $\sigma=0.5$ (compare to Figure~3).}
\label{fig:mdotscat}
\end{figure}

Figure~\ref{fig:LMscat} shows the {\it mean} relation between quasar luminosity
and halo mass in the presence of scatter, derived analogously to
Figure~\ref{fig:LM}. Figure~\ref{fig:mdotscat} shows the resulting accretion
rates, analogously to Figure~\ref{fig:mdot}.  Figures~\ref{fig:LMscat}
and~\ref{fig:mdotscat} are very similar to Figures~\ref{fig:LM}
and~\ref{fig:mdot}, respectively. In particular, a decline of $\dot m$ of
approximatively 2 orders of magnitude is still inferred between $0 \lsim z
\lsim 3$. Since, in first approximation, a given B-band luminosity is $\propto
M_{\rm bh} \dot M
\propto M_{\rm bh}^2 \dot m$ for thin disk emission (see
Fig.~\ref{fig:mdot}), a strong intrinsic scatter of $\sigma = 0.5$ for log
$M_{\rm bh}$ corresponds to a scatter $\simeq 2 \sigma$ for log $\dot
m$. Therefore, at the $1 \sigma$ level, an uncertainty of a factor $\sim 10$
exists on the inferred $\dot m$ shown in Figure~\ref{fig:mdotscat}. Even with
this large uncertainty, the decline of the accretion rate from high redshift to
low redshift is still statistically significant. We emphasize again here that
this amount of scatter in $\dot m$ corresponds to the extreme case where all
the observed scatter in the relation between $M_{\rm bh}$ and $M_{\rm bulge}$
is assumed to be intrinsic to the ratio $M_{\rm bh}/M_{\rm halo}$.

\section{X-ray vs. Optical Data}

A simultaneous study of the quasar evolution in X--ray and in optical ought to
reveal important information on accretion processes in these objects.  A
detailed treatment of this question, which would necessarily involve an X--ray
emission model, is beyond the scope of the present paper.  In general, the
ratio of optical to X--ray emission of quasars could be a complicated function
of $M_{\rm bh}$, $\dot m$ and $z$.  Quasars detected by ROSAT in the
soft X--rays
have revealed a different behaviour of the X--ray/optical ratio for
radio--quiet (Yuan et al. 1998) and radio--loud (Brinkmann et al. 1997)
objects. Radio-quiet quasars have in general steeper soft X-ray
spectra than radio-loud ones.  The X-ray loudness has been
found to be independent of redshift for both types of sources, but
a slight increase of the loudness has been found with
optical luminosity in the more abundant radio--quiet sources,
although this might be explained by the
dispersion of the intrinsic luminosities and the flux limits of
the observations.  The handful of $z>3$ ROSAT quasars confirm the
result that the X--ray/optical luminosity ratio is correlated with optical
luminosity and does not evolve strongly with redshift (Pickering et al. 1994).
On the other hand, recent ROSAT observations resolved the bulk of the
X--ray background into discrete sources (Hasinger 1999), and have shown that
the characteristic hard spectrum of the XRB can only be
explained if most QSO spectra are heavily absorbed.
This would imply the existence of a (possible redshift--dependent)
population of QSOs with very large X--ray/optical ratios.
A wealth of data exists on the X--ray/optical ratios in Galactic
BH candidates, and several authors have attempted to directly apply
these Galactic observations to supermassive BHs (see, e.g. Choi et al. 1999).

For simplicity, here we consider only the following, simplified question.  We
assume that the accretion rate previously derived from the optical LF is an
accurate estimate of $\dot m(z)$, and we expect the same accretion rate to
power X--ray emission from quasars. Therefore, we ask: what is the ratio of
optical to X--ray emission (as a function of $M_{\rm bh}$ and $z$) required to
obtain consistency between the accretion rate derived from the optical and the
X--ray luminosity functions.

\begin{figure}[t]
\includegraphics{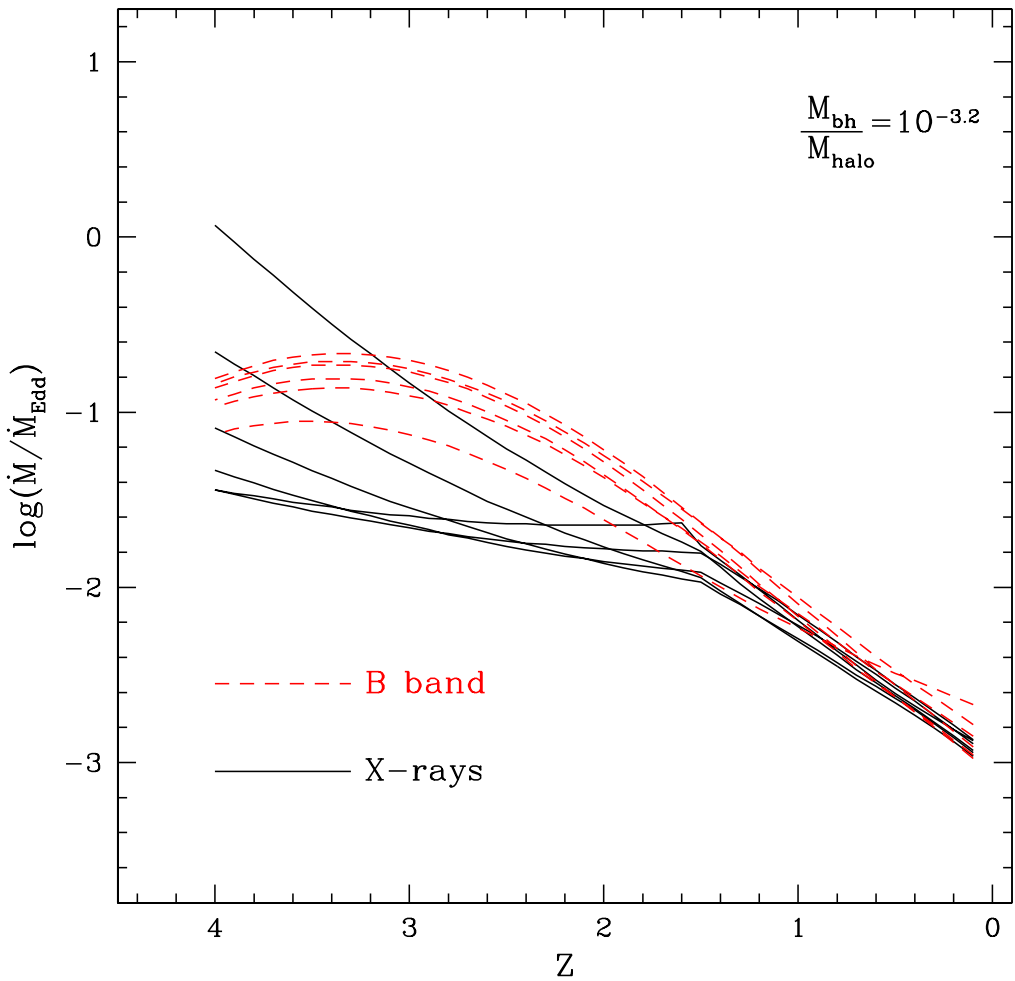}
\vspace*{4.5in}

\caption{
Mean accretion rates inferred from the quasar optical and X-ray LFs, and
Press-Schechter theory, assuming a {\it mean} ratio $M_{\rm bh}/M_{\rm
halo}=10^{-3.2}$ with an intrinsic scatter in $\log (M_{\rm bh}/M_{\rm halo})$
of $\sigma=0.5$, and $t_{\rm source}=10^7$~yr. The values of $\dot M$ inferred
from the optical data (dashed lines) correspond to the solid lines of
Fig.~\ref{fig:mdotscat}.  The values of $\dot M$ inferred from the X-ray data
(solid lines) have been derived assuming a ratio of optical to X-ray emission
which depends on $M_{\rm bh}$ (see text for details).  }

\label{fig:opticalvsX}
\end{figure}

In Figure~\ref{fig:opticalvsX}, the accretion rate derived from optical data,
assuming a strong scattering of $\sigma=0.5$ in the $\log$ of the ratio $M_{\rm
bh}/M_{\rm halo}$, with a mean of $10^{-3.2}$ and $t_{\rm source}=10^7$~yr, is
shown as dashed lines for various $M_{\rm bh}$ (equivalent to the solid lines
of Fig.~\ref{fig:mdotscat}). The solid lines in Figure~\ref{fig:opticalvsX}
represent the best match to the dashed lines of the $\dot m$ inferred from the
X--ray data when the same scattering, mean $M_{\rm bh}/M_{\rm halo}$ and
$t_{\rm source}$ are assumed. This match is obtained for the ratio of X--ray to
optical emission:

\begin{equation}
\label{eq:ratio}
\frac{\nu
L_{\nu}(X)}{\nu L_{\nu}(B)}=0.17 \left(
\frac{10^{10}~{\rm M_\odot}}{M_{\rm bh}} \right)^{0.4},
\end{equation}
where the B-band luminosity $\nu L_{\nu}(B)$, which is a function of $\dot m$,
is taken from the disk models described in \S~4.

Under these assumptions, the agreement between the $\dot m$ inferred from the
optical and X--ray LFs is excellent between $0 < z < 1.5$.  The agreement is
poor for $z>1.5$, where a significant dependence of $\dot m$ on $M_{\rm bh}$
appears (larger $\dot m$ being inferred for the more massive BHs).  We note,
however, that at redshifts above $z > 1.5$ the X--ray LF is not well
determined, and the fitting formula of Miyaji et al. (1998) is simply flat,
with no redshift evolution.  Therefore, although the results shown in
Figure~\ref{fig:opticalvsX} reveal a more complicated (although still
significantly declining) accretion rate at $z>1.5$, these results should be
considered more uncertain, until the X--ray LF is measured more accurately.
The environments of quasars can certainly be affected by cosmological structure
formation and evolution, which could lead to a redshift-evolution of the
X--ray/optical ratio.  In this case, the departure from equation~\ref{eq:ratio}
at high-$z$ could indeed be significant.

The above considerations reveal that the relation between the X--ray and the
optical emission of quasars, and its evolution, are not trivial. It suggests
that the energy radiated in X--rays is $\sim 20$ \% of the energy in the B-band
for a BH of mass $M_{\rm bh}=10^{10} M_{\odot}$, and higher for smaller BHs
(with approximate equality between optical and X--ray emission for $M_{\rm
bh}=10^8 M_{\odot}$).  The reason for the dependence on $M_{\rm bh}$ is
unclear. If one believes that at low enough $\dot m$ (say $< 0.1$), the
accretion flow in quasars is made of an inner ADAF and an outer thin disk, then
the increase of $\nu L_{\nu}(X)/\nu L_{\nu}(B)$ for smaller $M_{\rm bh}$ could
be due, at least in part, to an increase of the transition radius $R_{\rm tr}$
between the ADAF and the disk, which would result in a reduced optical emission
from the disk.  This trend seems consistent with the rather large values of
$R_{\rm tr}$ in X--ray binaries containing stellar-mass BHs (e.g.  Narayan,
Barret \& McClintock 1997) and the smaller value in, for example, the AGN NGC
4258 (Gammie, Narayan \& Blandford 1999).  However, the problem is complicated
by the fact that the value of $R_{\rm tr}$ also depends on $\dot m$ (see
e.g. Narayan et al. 1998 for a discussion).

The argument based on $R_{\rm tr}$ is only qualitative and tentative, and it
certainly does not capture all the physics of the problem. We note, for
example, that at high $z$ and nearly Eddington accretion rates, one cannot
invoke optically thin ADAFs to produce the large amount of X--ray emission
observed.  At smaller redshifts, the good agreement we find between the $\dot
m$ inferred from optical and X-ray data (using Eq.~\ref{eq:ratio}) implicitly
assumes the proportionality $\nu L_{\nu}(X)\propto \dot m$.  If the quasar's
emission were due to an ADAF alone, then the X--ray luminosity would scale
approximately as $\dot m^2$, rather than as $\dot m$ (e.g.  Narayan et
al. 1999).  Internal consistency of a two-component ADAF + thin disk model
therefore requires a dependence of $R_{\rm tr}$ on $\dot m(z)$ such that the
linear scaling $\nu L_{\nu}(X)\propto \dot m$ we assume is recovered.  While
this is plausible, a more detailed theoretical treatment of these models is
necessary to clarify whether ADAFs indeed play a role in the evolution of the
optical vs. X--ray LFs.

\section{Discussion}

In the class of models presented here, black holes grow synchronously
with their dark matter halos after their formation, maintaining a
constant ratio $M_{\rm bh}/M_{\rm halo}$.  We did not address the
important question of consistency between the mass accretion rates
inferred from the LFs and the individual black hole masses acquired by
the end of the quasar phase ($M_{\rm bh} = \int dt \dot M_{\rm bh}$).
This would require following the accretion and the merger history of
individual halos (Lacey \& Cole 1993; Kauffmann \& White 1993) while
our work concentrated on the quasar population as a whole.  Related to
this question is the interpretation of the duty--cycle, which could be
caused either by short intermittent activity phases or one single
luminous phase.  We assumed that all halos have a central black hole,
as suggested by observations (Magorrian et al. 1998); however the duty
cycle could be partly interpreted as a fraction of the halos harboring
quasars.

\begin{figure}[t]
\includegraphics{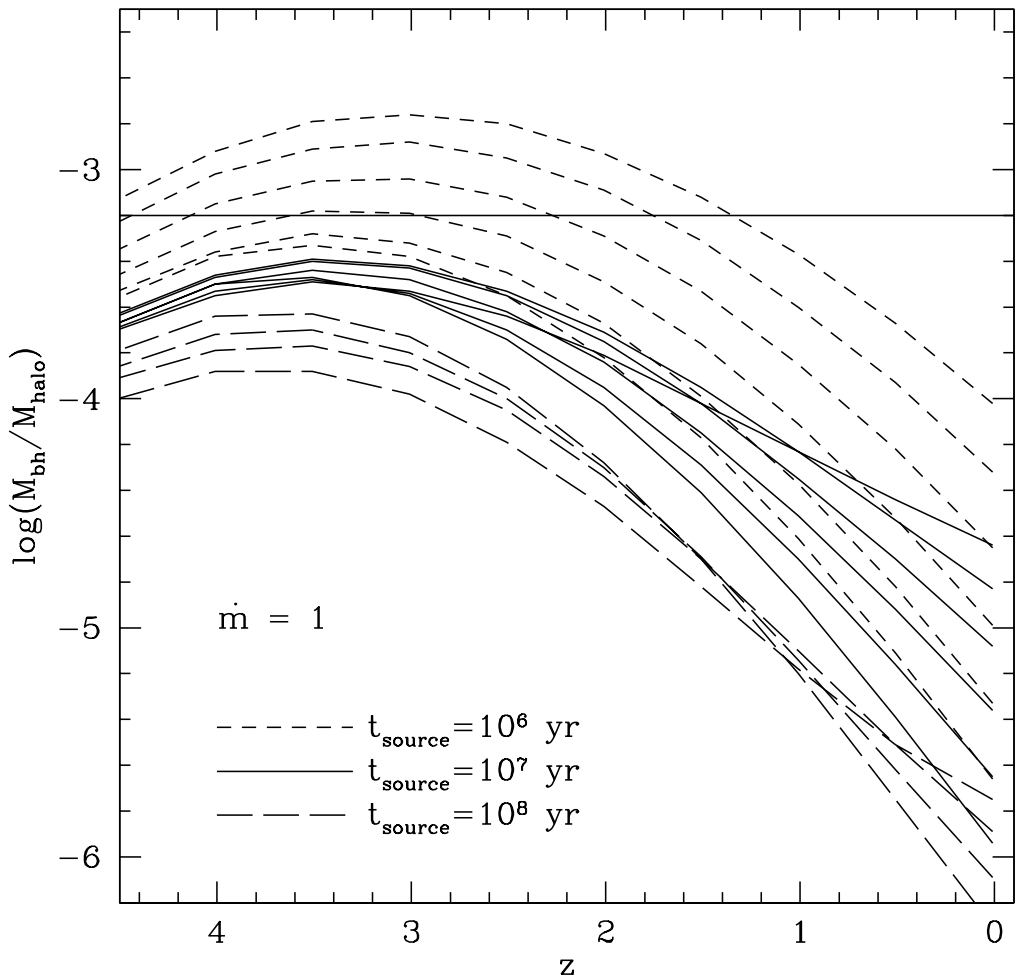}
\vspace*{4.5in}
\caption{The ratio $M_{\rm bh}/M_{\rm halo}$ as a function of redshift inferred
from the B--band LF, if a constant accretion rate $\dot{m}=1$ is assumed for
all redshifts.  The various curves correspond to halo masses in the range
$10^{11} \leq M_{\rm halo}/{\rm M_\odot} \leq 10^{14}$ (bottom to top at
$z=0$), and the results are shown for three different values of the quasar
lifetime, as in Figure~\ref{fig:mdot} and~\ref{fig:mdotscat}.  The horizontal
solid line shows the constant value of $M_{\rm bh}/M_{\rm halo}$ assumed in the
models with a declining accretion rate.}
\label{fig:mbhz}
\end{figure}

It is important to note that, independent of the questions of interpretation
above, our fit to the LF with a declining $\dot m(z)$ is not unique.  An
alternative possibility is a decrease of the average quasar lifetime with
redshift.  A continuously shortening lifetime would decrease the duty cycle,
and this would go in the right direction towards explaining the decline in the
QSO abundance at $z\lsim 3$.  Figure~\ref{fig:mdot} shows that keeping the
accretion rate constant ($\dot m=1$) over the range $0\lsim z\lsim 4$ would,
however, require very short lifetimes near $z\sim0$.
In order to quantify the role
$t_{\rm source}$ might play in the evolution of the LF, we have repeated our
calculations, keeping the accretion rate $\dot m$ fixed, and allowing $t_{\rm
source}$ to decrease with decreasing $z$.  We find that the lifetimes required
to fit the LF near $z\sim0$ are implausibly short, $t_{\rm source}\lsim$1 yr.
This is not surprising: reducing $\dot m$ allows a QSO at a fixed luminosity to
be associated with a more massive, and thus an exponentially rarer dark halo.
In comparison, a reduction in the duty--cycle decreases the abundance only
linearly with $t_{\rm source}$.  A second alternative is an accretion rate
kept constant at the Eddington value, while the ratio $M_{\rm bh}/M_{\rm halo}$
decreases rapidly towards low redshifts (Haehnelt, Natarajan \& Rees 1998).  In
Figure~\ref{fig:mbhz} we explicitly show the value of $M_{\rm bh}/M_{\rm halo}$
as a function of $z$ required in this case. In this model, the BH formation
efficiency decreases from redshift $z\approx3$ to $z\approx0$ by $\sim$2 orders
of magnitude.  The ratio $M_{\rm bh}/M_{\rm halo}$ is also a function of
$M_{\rm halo}$. The form of this function depends sensitively on the assumed
duty--cycle, but at $z=0$ the ratio is monotonically increasing with $M_{\rm
halo}$ for $t_{\rm source}\gsim10^6$ years.  In this model, at $z=0$, the
required BH masses are significantly smaller than considered above (\S~2); this
is consistent with the Magorrian et al. (1998) data, provided bulges comprise
only a small fraction of the total baryonic mass in galaxy--size halos, and the
bulge mass has a non--linear dependence on the halo mass (see Haehnelt,
Natarajan \& Rees 1998 for a discussion).

Observations of nearby galaxies provide only a relation between the
black hole mass and the mass of the spheroidal component of the galaxy
(Magorrian et al. 1998). It is possible that a significant fraction of
the baryonic mass of these galaxies (in form of gas) is omitted from
the mass budget when $M_{\rm bh}$ is related to $M_{\rm
halo}$. Indeed, attempts to observationally estimate the mass of an
extended gaseous component in both spiral and elliptical galaxies have
remained rather unsuccessful so far (see Zaritsky 1998). If this
component is important for the mass budget, halos of observed galaxies
are more massive than assumed above, which results in smaller values
of the (mean) ratio $x=M_{\rm bh}/M_{\rm halo}$. The effect of
reducing $x$ by a constant factor is simply to scale up the accretion
rates required to fit the quasar LF, but the decline with $z$ is
conserved.  For example, for the case with strong intrinsic scatter,
we find that the effect of reducing the mean value of the ratio $x$
from $10^{-3.2}$ to $10^{-4.2}$ is to scale up $\dot m(z)$ by one to
two orders of magnitude (depending on halo mass and $t_{\rm source}$)
in comparison to the values shown in Figure~\ref{fig:mdotscat}. This
is unacceptable in the framework of the thin accretion disk models
considered here since the accretion rates at high $z$ become
substantially super-Eddington for all halo masses and $t_{\rm source}
\gsim 10^6$ yr.

In deriving the relations $L-M_{\rm halo}$ shown in Figures~\ref{fig:LM}
and~\ref{fig:LMscat}, only two assumptions have been made: (1) one BH in each
Press--Schechter halo, and (2) a constant quasar lifetime.  In particular, the
$L-M_{\rm halo}$ relation is independent of the ratio $M_{\rm bh}/M_{\rm
halo}$, and quite large halo masses ($ 10^{13}~{\rm M_\odot} \lsim M_{\rm halo}
\lsim 10^{15}~{\rm M_\odot}$) arise simply by equating the predicted space
densities of halos with the observed space densities of optical quasars (a
similar result is obtained with the X-ray LF). There are two ways to avoid
unreasonably large halo masses.  The first option is to have a duty cycle
$t_{\rm source}\ll 10^6$ yr. Such a short $t_{\rm source}$ could be consistent
with super-Eddington accretion via a ``radiation torus'' (Blandford \& Rees
1992), but not with the Eddington--limited thin disk accretion considered here.

A second, perhaps more attractive, solution to the problem of large halo masses
is to observe that such halos ($M_{\rm halo} \gsim 10^{13}~{\rm M_\odot}$)
correspond to groups or clusters of galaxies in the Press--Schechter formalism.
We have so far assumed that each halo harbors one BH (as did previous works
relating quasars to Press--Schechter halos).  However, it is plausible that a
cluster--sized halo contains several quasars, perhaps one in each of its member
galaxies.  Including this effect would change the black hole mass function, and
result in a reduction of $M_{\rm bh}$ and an increase of $\dot M$ deduced from
the observations.  A detailed study of this effect requires a knowledge of the
sub--structure of each halo, which is beyond the scope of the present work.

Although we based our models on the spectrum of a thin accretion disk, the
decline in the accretion rate would be inferred with more general accretion
models as well. The two essential characteristics of thin disk accretion that
we utilized are (1) a radiative efficiency of roughly $10 \%$, and (2) the
approximate proportionality of the flux to $M_{\rm bh}$ and $\dot m$. Any
accretion model with roughly these properties would lead to similar conclusions
on the decline of the accretion rate with decreasing redshift.
Finally, in order to assess the sensitivity of our results
to the assumed cosmology, we repeated our calculations in an open
universe (setting $\Lambda=0$).
The predicted space density from the Press--Schechter formalism
is strongly dependent on cosmology.  However, when we
assume a different cosmology, we also correct the observationally
derived luminosity function,  since this
quantity comes from the number of objects on the sky per
solid angle.  Using the corrected
intrinsic space density $dN_{\rm obs}/dL$ tends to balance the changes
to the Press--Schechter mass function due to cosmology.
We find that the accretion rates inferred
in the open CDM model differ by less than $\sim 20\%$ from those
shown in Figure~\ref{fig:mdot} in the $\Lambda$CDM model.

\section{Conclusions}

We constructed models for the cosmological evolution of quasars, using
the Press--Schechter theory for determining the black hole mass
function, and assuming that quasar optical emission is due to
accretion via a thin disk. Contrary to existing models of quasar
evolution, the accretion rate $\dot M$ of the quasar population is not
postulated, but rather inferred from the observed quasar luminosity
function.

According to these models, the accretion rate of the population of quasar 
black holes decreases with cosmic time. 
We find that the same fueling rate is inferred from the X--ray and 
optical LF, provided the X--ray/optical flux ratio decreases with 
BH mass.  More detailed modeling of the emission processes in these
two bands is needed to clarify the validity of the relation we find.
A peak in the accretion rate near $z\simeq
3.5$ (Fig. 3) is inferred from the optical LF.  A derivation using X--ray data
does not show a similar turnover, suggesting that the peak inferred from the
optical data is caused by dust obscuration. Near $z \sim 0$, $\dot M$ drops to
substantially sub-Eddington values at which ADAFs exist. The combination of a
decreasing $\dot M(z)$ and a possible transition to radiatively inefficient
ADAFs at late times could be the origin of the absence of bright quasars in the
local universe {\it and} the faintness of accreting BHs at the centers of
nearby galaxies. We argued that such a decline of the quasar accretion rate
over cosmic timescale is consistent with expectations from cosmological
structure formation models, and cannot be explained by halo mergers alone if
the fueling rate of quasars is proportional to the merger rate.

\acknowledgments

We thank L. Hernquist, A. Loeb and R. Narayan for useful discussions
and comments.  ZH was supported by the DOE and the NASA grant NAG
5-7092 at Fermilab. KM was supported by a SAO Predoctoral Fellowship,
NSF grant 9820686, and a French Higher Education Ministry grant.

\clearpage
\newpage

\clearpage
\newpage

\clearpage
\newpage

\clearpage
\newpage

\clearpage
\newpage

\clearpage
\newpage

\clearpage
\newpage

\clearpage
\newpage

\clearpage
\newpage

\end{document}